\documentclass[journal,twoside]{IEEEtran}

\let\labelindent\relax

\usepackage{cite}
\usepackage{graphicx}
\usepackage{psfrag}
\usepackage{amsmath}
\usepackage{multirow}
\usepackage{amssymb}
\usepackage{amsthm}
\usepackage{mathrsfs}
\usepackage{array}
\usepackage{color}
\usepackage{booktabs}
\usepackage{lipsum}
\usepackage{threeparttable}
\usepackage{booktabs}
\usepackage{makecell}
\usepackage{tabularx}
\usepackage{subcaption}
\usepackage[font=small,labelfont=bf]{caption}
\usepackage{enumitem,kantlipsum}

\begin{document}
\title{
	\LARGE{An Enhanced Random Access with Preamble-Assisted Short-Packet Transmissions for Cellular IoT Communications\\ \large{(Extended Version)$^\ast$}}
	\author{Taehoon Kim,~\IEEEmembership{Member,~IEEE,} and Inkyu Bang,~\IEEEmembership{Member,~IEEE}}
	\thanks{$^\ast$This is an extended version of the journal paper accepted to be published in IEEE Communications Letter with the same title. This is the authors$'$ version of the work. It is posted here for your personal use. Not for redistribution.}
	\thanks{T. Kim and I. Bang$^\dagger$ are with the Agency for Defense Development, Daejeon, 34186, Republic of Korea (e-mail: \{taehoonkim,ikbang\}@add.re.kr).}
	\thanks{$^\dagger$\emph{Corresponding author} and research done while working at National University of Singapore. This work was supported by the National Research Foundation of Korea (NRF-2018R1A6A3A03012996).}
}
%\markboth{IEEE Communications Letters, VOL. X, NO. X, XX, 2019}{Kim \MakeLowercase{\emph{et al.}}: An Enhanced Random Access with Preamble-Assisted Short-Packet Transmissions for Cellular IoT Communications}
	
\maketitle

\begin{abstract}
	We propose an enhanced random access (RA) with preamble-assisted short-packet transmissions to support cellular Internet-of-things (IoT) communications. 
	A key feature of the proposed scheme is that the base station (e.g., eNodeB in LTE networks) utilizes the RA preambles as uplink reference signals to estimate uplink channel state information of IoT devices, which subsequently provides additional opportunities to transmit short-packets during the RA procedure without extra signaling (e.g., connection and scheduling requests). Through simulations, we evaluate the performance of the proposed scheme in terms of a channel estimation mean squared error, a bit error rate, a preamble collision probability, and a success probability of short-packet transmissions. The results show that the proposed scheme can support reliable and low-latency featured short-packet transmissions during the RA procedure by efficiently incorporating both multiple antenna and detection techniques.
\end{abstract}


\section{Introduction}
\IEEEPARstart{W}{ith} emerging information technologies, the idea of {Internet-of-things (IoT)} has proliferated to support our convenient life~\cite{Palattella2016}. In a wide range of IoT applications such as smart metering and e-health, most IoT devices are expected to generate small-sized packets rather than long-sized ones, which is a major difference from the design goal of the cellular networks (e.g., LTE/LTE-A). Accordingly, there have been a number of studies to tailor the commercial cellular networks to support the traffic generated by the IoT devices~\cite{TR_36_881}.

A connection-oriented scheduling mechanism in cellular systems makes a challenge in supporting short-packet transmissions. For every data transmission, each device should request uplink resources after establishing a connection through a random access (RA) procedure. Thus, the signaling overhead (i.e., a ratio of the number of bits to the amount of signalings) significantly increases as the packet size decreases~\cite{al2015internet}.

A number of studies have investigated to efficiently support short-packet transmissions in cellular networks~\cite{Dai2018,Wiriaatmadja2015,Jang2016,Kim2017}. Grant-free protocols have been studied in recent~\cite{Dai2018} where each IoT device transmits short-packets through the pre-reserved radio resources instead of on-demand connection and scheduling requests. This approach may effectively reduce the signaling overhead during the short-packet transmissions. However, it requires significant modifications in the current cellular networks. To overcome this limitation, data transmission during the RA procedure was investigated. Wiriaatmadja and Choi~\cite{Wiriaatmadja2015} and Jang \emph{et al.}~\cite{Jang2016} proposed a hybrid RA scheme and a message-embedded RA scheme, respectively. However, the performance of the previous work is still limited due to the collision problem during the contention-based RA procedure.

Recently, Kim \emph{et al.}~\cite{Kim2017} proposed a novel RA preamble detector which utilizes multiple Zadoff-Chu (ZC) sequences. This showed the feasibility of an almost collision-free environment by increasing the number of RA preambles without any performance degradation in the RA procedure.\footnote{As the number of ZC sequences increases, the number of available preambles increases, which is effective to alleviate preamble collisions. However, the occurrence of erroneous detection increases due to the noise-rise among ZC sequences with different root indices.} 
Further, we notice that the RA preambles may replace the role of uplink reference signals when the preamble collision problem is effectively alleviated and, thus, the data packets can be correctly decoded only with the channel estimates using the RA preambles. This motivates us to propose a notion of \emph{preamble-assisted short-packet transmissions}, which is compatible with the RA procedure without significant modifications.

In this letter, we propose an enhanced RA scheme to support short-packet transmissions. The proposed scheme additionally utilizes RA preambles to estimate the uplink channel state information of IoT devices, and enables the IoT devices to transmit short-packets along with the RA procedure without any extra signalings. Simulation results show that only a few more root sequences are sufficient to newly provide an additional functionality of reliable and low-latency featured short-packet transmissions during the RA procedure. 

\section{System Model}
\begin{figure}[t]
	\centering
	\includegraphics[width=8.5cm]{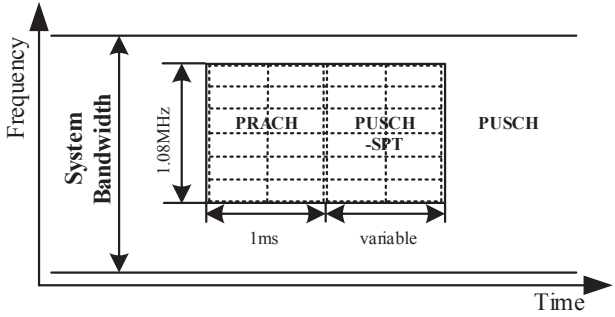}
	\caption{Uplink radio frame structure}
	\label{fig:Framework}
\end{figure}

We consider a single LTE/LTE-A cell network, where the eNodeB is equipped with $M$ antennas. Each IoT device with a single antenna attempts its RA at the {next available} RA slot, periodically configured in time domain, when {a new packet is ready for transmissions}. Here, {we consider a single RA slot}, where $N_\mathrm{I}$ devices simultaneously attempt RAs.\footnote{Considering a Poisson arrival model, {$N_\mathrm{I} = \frac{N \lambda  T_\mathrm{P}}{1-p_c}$}~\cite{Kim2015}, where $N$, $\lambda$, $T_\mathrm{P}$, and $p_c$ denote the number of IoT devices within a cell, a packet arrival rate, a period of RA slot, and a collision probability, respectively. {As $p_c$ increases, collision events cause an increasing of the load per RA slot due to backlogging~\cite{Laya2014} in the upcoming slots. Especially, when $N_\mathrm{I}\ge -1/\mathrm{ln}\left(1-\frac{1}{KN_\mathrm{P}}\right)$, the throughput of PRACH starts to be deteriorated.}}
The eNodeB configures $K$ ZC sequences as root sequences to generate RA preambles. Let $N_\mathrm{P}$ denote {the number of RA preambles generated from} a single ZC sequence. 
We consider a Pedestrian B channel model to represent the wireless channel environment in fixed/low-mobility IoT scenarios\cite{b_LTE}. We assume an open-loop power control of the IoT device, which guarantees a certain level of received signal-to-noise ratio (SNR) at the eNodeB. 
For short-packet transmissions, we consider an uncoded binary phase shift keying (BPSK) modulation and each modulated symbol is carried by each subcarrier in the physical uplink shared channel (PUSCH), reserved for short-packet transmissions, so-called PUSCH-SPT. Each device fully utilizes the entire resources within the PUSCH-SPT, and the eNodeB uses a zero-forcing (ZF) decoder for decoding of short-packets received via PUSCH-SPT.
To be specific, the PUSCH-SPT is located right next to the physical random access channel (PRACH) as shown in Fig.~\ref{fig:Framework}. In stationary environments, channel coherence time is usually longer than LTE frame (i.e., 10~ms). Thus, channel state information of the PRACH and that of the PUSCH-SPT are almost the same. The sub-carrier spacing of the PUSCH-SPT is assumed to be the same with that of the PRACH (i.e., $1.25$ kHz). Note that the time duration of the PUSCH-SPT has to be carefully set based on a given channel coherence time. %For analytical tractability, we set this value as $1$~ms. 
%%%%%%%%%%%%%%%%%%%%%%%%%%%%%%%%%%%%%%%%%%%%%%%%%%%%%%%%%%%%%%%%%%%%%
%%%%%%%%%%%%%%%%%%%%%%%%%%%%%%%%%%%%%%%%%%%%%%%%%%%%%%%%%%%%%%%%%%%%%

\section{An Enhanced Random Access for Preamble-Assisted Short-Packet Transmissions}
ZC sequence has been used to generate not only RA preamble but also demodulation reference signal (DM-RS) in LTE~\cite{b_LTE}. Thus, the RA preamble might be feasible to replace the role of the uplink reference signal.
%ZC sequence has been used to generate not only RA preamble but also demodulation reference signal (DM-RS) in LTE systems~\cite{b_LTE}. This fact motivates us to investigate the feasibility that the RA preamble can replace the role of uplink reference signal.
However, RA preambles cannot be directly used to estimate the uplink channel state information (CSI) due to preamble collisions during the RA procedure, which consequently results in wrong channel estimations. 
If this functionality is available with the RA procedure, preamble-assisted short-packet transmissions are feasible and able to effectively reduce signaling overhead during the transmissions. 
%If this functionality is envisioned along with the RA procedure, preamble-assisted short-packet transmissions become feasible and contribute to effectively reduce signaling overhead during the short-packet transmissions. 
In this section, we propose an enhanced RA procedure to support short-packet transmissions, and its main features are summarized as follows:

\begin{itemize}
	\item The proposed scheme utilizes multiple root sequences to sufficiently increase the number of available RA preambles, which effectively mitigates preamble collisions.
	
	\item The proposed scheme enables the eNodeB to estimate the uplink CSI of IoT devices using the RA preambles and decode short-packets using the estimated CSI.
\end{itemize}

\subsection{Procedure}
In detail, the proposed scheme consists of four steps as follows:
\begin{enumerate}[leftmargin=*,listparindent=1.0em]
	\item \textbf{Preamble and short-packet transmissions}: 
	Each device generates an RA preamble based on randomly selected root and preamble indices, $r \in \left\{r_1,...,r_K\right\}$, and $p \in \left\{1,...,N_\mathrm{P}\right\}$, respectively.
	The generated RA preamble is expressed as $x[n] = z_r[(n+p\cdot N_\mathrm{cs})_{\bmod}N_\mathrm{zc}]$ for $n=0,..,N_\mathrm{zc}-1$, where $z_r[n]$, $N_\mathrm{cs}$, and $N_\mathrm{zc}$ denote the ZC sequence with a root index of $r$, the size of cyclic shift, and the length of the ZC sequence, respectively~\cite{b_LTE}. After preamble transmissions through the PRACH, each device transmits a short-packet through the PUSCH-SPT. Note that any mapping rule between the selected preamble and the location of resources within the PUSCH-SPT is not required since each device utilizes the entire resources within the PUSCH-SPT.

	\item \textbf{Channel estimation through RA preambles}: 
	The eNodeB estimates the uplink CSI through the received RA preambles. This step should be performed for each of received signals from the entire antennas, i.e., $y^m[n]$ for all $m$, but we describe it from the perspective of $y^m[n]$ to clearly explain the key feature. Let $y^m[n]$ denote the time-domain received signal through the PRACH at the antenna $m$, 
	\begin{equation}
		\label{eq:rs_multi_root}
		y^m [n] = \sum\limits_{i \in \left\{ 1, \dots, N_\mathrm{I} \right\} } {\sqrt \beta \cdot h_{m,i} [n] \otimes x_i [n]}  + w_m [n],
	\end{equation}
	where $\beta$ denotes the signal strength, $h_{m,i}[n]$ denotes the channel coefficient from the IoT device $i$ to the antenna $m$ of the eNodeB, $x_i[n]$ denotes the transmitted RA preamble from the IoT device $i$, $w_m[n]$ denotes the additive Gaussian noise at the antenna $m$ of the eNodeB with zero mean and variance of $\sigma^2$, and $\otimes$ denotes a convolution operation. 	
	
	Fig.~\ref{fig:BlockDiagram} shows the channel estimation procedure. The correlation between ZC sequences with the same root index shows an ideal auto-correlation property (i.e., $\sum\nolimits_{n = 0}^{N_\mathrm{zc}  - 1} {z_{r_k } [n] \cdot z_{r_k }^* [n + \alpha ]}=\delta(\alpha)$) and it is useful to explicitly grasp multipath channels that each of RA preambles experiences.	To acquire channel estimates, the eNodeB calculates correlations between $y^m$ and the entire root sequences (i.e., $\left\{z_{r_1},...,z_{r_K}\right\}$). The correlation at lag $\alpha$ between $y^m$ and $z_{r_k}$ is denoted by $c_{ y^m ,z_{r_k } } [\alpha ]$ for $\alpha = 0,..,N_\mathrm{zc}-1$ in~\eqref{eq:correlation} shown at the top of the next page, where $\mathcal{I}_k$ and $(\cdot)^*$ denote a set of IoT devices choosing root index $k$ and a conjugate operation, respectively. 
	Accordingly, multiple channels between each IoT device and the eNodeB are captured from the correlation result. 	
	%Accordingly, multiple channels between each of IoT devices and the eNodeB are captured from the correlation result. 
	Thus, the eNodeB should separate those of channel impulse responses (CIRs) to clearly differentiate the channel estimates~\cite{Li2014}.
	\begin{table*}
		\begin{equation}
		\begin{aligned}
%		c_{ y^m ,z_{r_k } } [\alpha ] &= \sum\limits_{n = 0}^{N_\mathrm{zc}  - 1} {y^m [n] \cdot z_{r_k }^* [n+\alpha]} = \underbrace {\sum\limits_{n = 0}^{N_\mathrm{zc}  - 1} {\left( {\sum\limits_{i_k \in \mathcal{I}_k} {\sqrt \beta \cdot h_{m,i_k} [n] \otimes } z_{r_k } \left[ {\left( {n + p_{i_k}  \cdot N_\mathrm{cs} } \right)_{\bmod } N_\mathrm{zc} } \right]} \right) \cdot z_{r_k }^* [n+\alpha]}}_{\text{Desired signal from the viewpoint of } z_{r_k}}  \\	
%		& \quad\quad + \underbrace {\sum\limits_{n = 0}^{N_\mathrm{zc}  - 1} {\left( \sum\limits_{k'\in K, k'\neq k} {\sum\limits_{i_{k'} \in \mathcal{I}_{k'}} {\sqrt \beta \cdot h_{m,i_{k'}} [n] \otimes } z_{r_{k'} } \left[ {\left( {n + p_{i_{k'}}  \cdot N_\mathrm{cs} } \right)_{\bmod } N_\mathrm{zc} } \right]} \right) \cdot z_{r_k }^* [n+\alpha]} }_{\text{Intra-cell interference signal from the viewpoint of } z_{r_k}} + 	\sum\limits_{n = 0}^{N_\mathrm{zc}  - 1} {w_m [n] \cdot z_{r_k }^* [n + \alpha ]} 
		c_{ y^m ,z_{r_k } } [\alpha ] &= \sum\limits_{n = 0}^{N_\mathrm{zc}  - 1} {y^m [n] \cdot z_{r_k }^* [n+\alpha]} \\ 
		&= \sum\limits_{n = 0}^{N_\mathrm{zc}  - 1} {  \left(  \sum\limits_{i \in \left\{1,..,N_\mathrm{I}\right\}} {\sqrt \beta \cdot h_{m,i} [n] \otimes x_i [n]}  + w_m [n] \right) \cdot z_{r_k }^* [n+\alpha]} \\
		&= \underbrace {\sum\limits_{n = 0}^{N_\mathrm{zc}  - 1} {\left( {\sum\limits_{i_k \in \mathcal{I}_k} {\sqrt \beta \cdot h_{m,i_k} [n] \otimes } z_{r_k } \left[ {\left( {n + p_{i_k}  \cdot N_\mathrm{cs} } \right)_{\bmod } N_\mathrm{zc} } \right]} \right) \cdot z_{r_k }^* [n+\alpha]}}_{\text{Desired signal from the viewpoint of } z_{r_k}}  \\	
		& \quad\quad + \underbrace {\sum\limits_{n = 0}^{N_\mathrm{zc}  - 1} {\left( \sum\limits_{k'\in K, k'\neq k} {\sum\limits_{i_{k'} \in \mathcal{I}_{k'}} {\sqrt \beta \cdot h_{m,i_{k'}} [n] \otimes } z_{r_{k'} } \left[ {\left( {n + p_{i_{k'}}  \cdot N_\mathrm{cs} } \right)_{\bmod } N_\mathrm{zc} } \right]} \right) \cdot z_{r_k }^* [n+\alpha]} }_{\text{Interference signal from the viewpoint of } z_{r_k}} + 	\sum\limits_{n = 0}^{N_\mathrm{zc}  - 1} {w_m [n] \cdot z_{r_k }^* [n + \alpha ]} 
		\label{eq:correlation} 
		\end{aligned}
		\end{equation}	
		\hrulefill
	\end{table*}
	
	Especially, when the eNodeB configures multiple ZC sequences, the orthogonality among the RA preambles does not hold any more due to the cross-correlation property between ZC sequences with different root indices (i.e., $\sum\nolimits_{n = 0}^{N_\mathrm{zc}  - 1} {z_{r_k } [n] \cdot z_{r_{k'} }^* [n + \alpha ]}\neq\delta(\alpha) \text{ if } k\neq k'$). Thus, it is difficult to acquire the correct CSI from the received RA preambles due to the interference among RA preambles generated from different root sequences as shown in~\eqref{eq:correlation}.
	
	To resolve this problem, the received signal should be processed in advance. We thus employ an enhanced PRACH detector proposed in~\cite{Kim2017}, which effectively separates $y^m$ into multiple signals with different root sequences (i.e., $y^m \to \left\{ {\hat y_{r_1 }^m ,.., \hat y_{r_K }^m } \right\}$).\footnote{Even though we consider a single cell scenario, the same principle can be applied to the multi-cell scenario. Preliminary knowledge to the detector is a set of root sequences utilized at the home cell and the neighboring cells. This pre-processing effectively alleviates the interference caused by other root sequences but $\hat y_{r_k}^m$ may still contain signal reconstruction errors.} Consequently, the eNodeB acquires the channel estimates using $c_{ \hat y_{r_k}^m ,z_{r_k } } [\alpha ]$ for all $k$ instead of $c_{ \hat y^m ,z_{r_k } } [\alpha ]$ for all $k$. Note that the complexity of the channel estimation procedure depends on that of the PRACH detector~\cite{Kim2017}, which is linearly increased according to the number of used ZC sequences. But the complexity is not high enough to be handled at the eNodeB.

	 Fig.~\ref{fig:Estimation_Results} shows an example of channel estimations when three IoT devices are considered. Orthogonal preambles can provide accurate channel estimates but the preamble collision may occur due to the limited number of available RA preambles, which results in completely wrong channel estimates. Non-orthogonal preambles effectively mitigate the collision problem while slightly degrading the accuracy of channel estimation.
	
	%%%%%%%%%%%%%%%%%%%%%%%%%%%%%%%%%%%%%%%%%%%%%%%%%%%%%%%%%%%%%%%%%%%%%
	\begin{figure}[t]
		\centering
		\includegraphics[width=8.9cm]{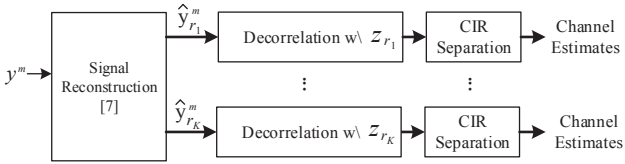}
		\caption{Channel estimation procedure}
		\label{fig:BlockDiagram}
	\end{figure}
	%%%%%%%%%%%%%%%%%%%%%%%%%%%%%%%%%%%%%%%%%%%%%%%%%%%%%%%%%%%%%%%%%%%%%	
	
	%%%%%%%%%%%%%%%%%%%%%%%%%%%%%%%%%%%%%%%%%%%%%%%%%%%%%%%%%%%%%%%%%%
	\begin{figure}[t]
		\centering
		\begin{subfigure}[t]{0.5\textwidth}
			\centering
			\includegraphics[width=8.7cm]{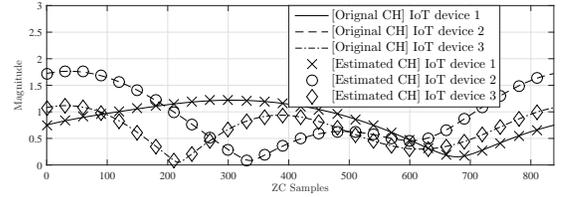}
			\caption{Channel estimates with orthogonal preambles (no collision)}
			\vspace{3.0mm}
		\end{subfigure}

		\begin{subfigure}[t]{0.5\textwidth}
			\centering
			\includegraphics[width=8.7cm]{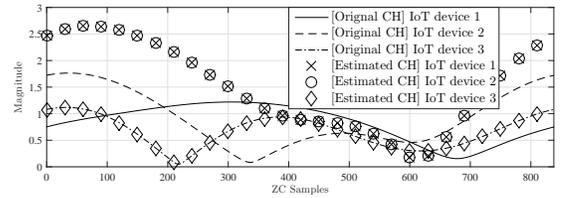}
			\caption{Channel estimates with orthogonal preambles (collision)}
			\vspace{3.0mm}
		\end{subfigure}

		\begin{subfigure}[t]{0.5\textwidth}
			\centering
			\includegraphics[width=8.7cm]{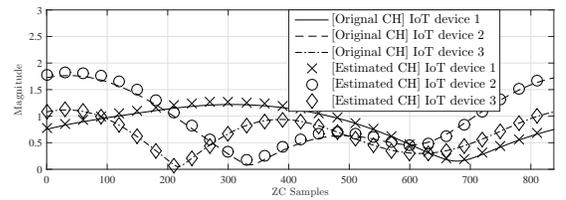}
			\caption{Channel estimates with non-orthogonal preambles (no collision)}
		\end{subfigure}
		\caption{Comparison of channel estimates when $N_\mathrm{I}=3$}
		\label{fig:Estimation_Results}
	\end{figure}
	%%%%%%%%%%%%%%%%%%%%%%%%%%%%%%%%%%%%%%%%%%%%%%%%%%%%%%%%%%%%%%%%%%%%%	
	
	\item \textbf{Data decoding}: Using the channel estimates (i.e., $\hat h_{m,i},~\forall m, \forall i$) acquired in step (2), the eNodeB decodes short-packets received through the PUSCH-SPT based on the ZF decoder. 
	When RA preambles are received without any collisions and {mis-/erroneous detections}, the full rank ZF matrix (i.e., the number of independent RA preambles) to successfully decode short-packets is available. Otherwise, the ZF matrix does not satisfy the full rank condition, which results in decoding failure.
	
	\item \textbf{Acknowledgment}: The eNodeB transmits the acknowledgment to successfully decoded short-packets.
\end{enumerate}

\section{Numerical Results}
We perform link level simulations using LTE/LTE-A related parameters specified in Table~\ref{tb:simulation_env}.
%%%%%%%%%%%%%%%%%%%%%%%%%%%%%%%%%%%%%%%%%%%%%%%%%%%%%%%%%%
\begin{table}[t]
	\begin{center}
		\caption{Simulation parameters and values}		
		\begin{tabular}{l|l}
			\toprule
			Parameters & Values  \\
			\midrule
			Number of antennas at the eNodeB ($M$) & 8, 16 \\
			Number of IoT devices ($N_\mathrm{I}$) & 1 $\sim$ 8 \\
			Number of root sequences ($K$) & 1, 5 \\
			SNR ($\beta / \sigma^2$) & 5 $\sim$ 25 dB\\
			$N_\mathrm{zc}$, $N_\mathrm{cs}$, $N_\mathrm{P}$ & 839, 13, 64 \\
			\bottomrule
		\end{tabular}
		\label{tb:simulation_env}
	\end{center}
\end{table}
%%%%%%%%%%%%%%%%%%%%%%%%%%%%%%%%%%%%%%%%%%%%%%%%%%%%%%%%%%
We also consider the following metrics to evaluate the proposed scheme.
%We also consider the following metrics to measure the performance of the proposed scheme.

\begin{enumerate}[leftmargin=*,listparindent=1.0em]
	\item \textbf{Mean squared error (MSE)}: MSE quantifies the accuracy of channel estimation. When the channel estimates differ from the original channel realizations, the data can be wrongly decoded. The MSE given by $\eta$ is calculated as 
	\begin{equation}
	\eta = \mathbb{E}\left[\left(h_{m,i} - \hat h_{m,i} \right) \left( h_{m,i} - \hat h_{m,i} \right)^*\right].
	\end{equation}
	Note that a closed-form derivation of the channel estimate $\hat h_{m,i}$ is difficult in general due to the error propagation effects during the signal reconstruction, which also implies that the accuracy of $\hat h_{m,i}$ decreases as $K$ increases.
	
	\item \textbf{Bit error rate (BER)}: BER quantifies the reliability of the transmissions, denoted by $p_b\triangleq \Pr\left\{s_\mathrm{tx}\ne s_\mathrm{rx} \right\}$, where $s_\mathrm{tx}$ and $s_\mathrm{rx}$ represent the transmitted and received bit, respectively. Since the accuracy of $\hat h_{m,i}$ is the most critical factor to affect $p_b$, $p_b$ also decreases as $K$ increases.
	
	\item \textbf{Collision probability}: Preamble collision occurs when two or more devices select the same root and preamble indices. When the number of RA attempting IoT devices at a certain RA slot is given by $N_\mathrm{I}$, the collision probability denoted by $p_c$ can be expressed as 
	\begin{equation}
	p_c = 1 - \left(1 - \frac{1}{(K\cdot N_\mathrm{P}) } \right)^{N_\mathrm{I}-1},
	\label{eq:collision_prob}
	\end{equation} 
	which decreases as $K$ increases. 
	
	\item \textbf{Success probability}: Short-packets are successfully transmitted when each IoT device does not experience any preamble collisions and transmission errors.\footnote{We do not take mis- and erroneous detections into consideration, since their occurrences can be made negligible by adjusting the detection threshold.} Thus, a success probability of short-packet transmissions is defined as
	\begin{align}
	p_s  = \left\{ \begin{array}{l}
	(1 - p_c )(1 - p_b)\approx 1-p_c,~\mathrm{if}~N_\mathrm{I}\le M \\ 
	0,~\mathrm{otherwise}. \\ 
	\end{array} \right.
	\label{eq:success_prob}
	\end{align}
	Note that $p_s$ mainly depends on $p_c$ when a reliable communication link is guaranteed (i.e., $p_b < 10^{-3}$).
	Increasing $K$ slightly degrades $p_b$ due to the inaccuracy of channel estimations caused by the signal reconstruction errors but it can significantly decrease $p_c$ in~\eqref{eq:collision_prob}. Thus, increasing $K$ can effectively improve $p_s$.
\end{enumerate}

\begin{figure}[t]
	\centering
	\includegraphics[width=8.7cm]{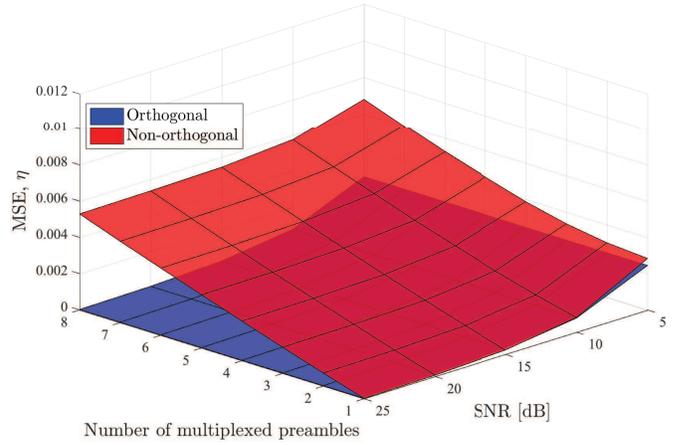}
	\caption{MSE for varying both the number of multiplexed preambles and SNR value}
	\label{fig:Fig5}
\end{figure}

Fig.~\ref{fig:Fig5} shows the MSE between the channel realizations and estimates for varying the number of multiplexed preambles and SNR when the proposed scheme is applied.  The MSE highly depends on the number of multiplexed preambles. When orthogonal preambles are considered, the MSE shows a constant value. On the contrary, when non-orthogonal preambles are considered, the MSE slightly increases due to the error propagation effect during the signal reconstruction. Despite a slight loss in accuracy of channel estimations, the result reveals the feasibility of the proposed scheme to be used in practice. {Note that the channel estimation is \emph{independently} performed for the received signals from each of antennas, and, thus, the number of antennas does not affect the MSE performance.}
%%%%%%%%%%%%%%%%%%%%%%%%%%%%%%%%%%%%%%%%%%%%%%%%%%%%%%%%%%%%%%%%%%%%%
%%%%%%%%%%%%%%%%%%%%%%%%%%%%%%%%%%%%%%%%%%%%%%%%%%%%%%%%%%%%%%%%%%%%%

\begin{figure}[t]
	\centering
	\begin{subfigure}{0.24\textwidth}
		\includegraphics[width=1.75in]{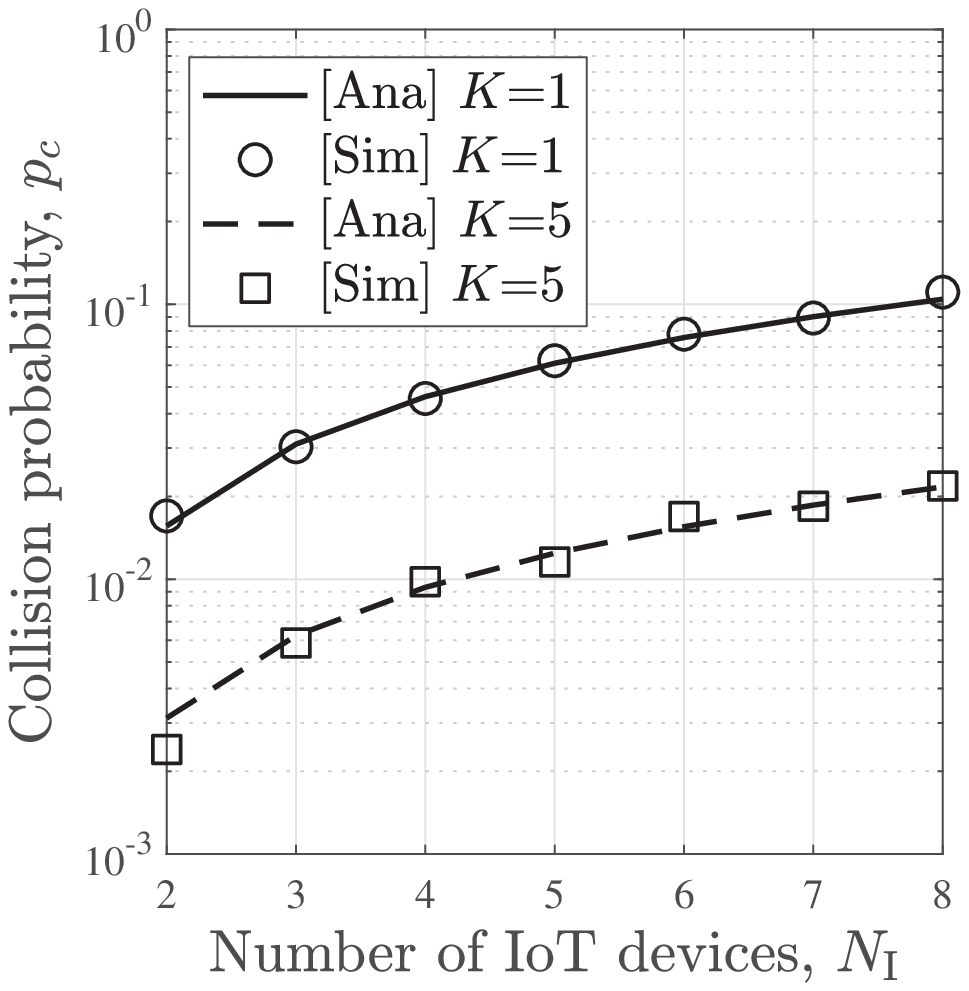}
		\caption{Collision probability}
	\end{subfigure}	
	\begin{subfigure}{0.24\textwidth}
		\includegraphics[width=1.75in]{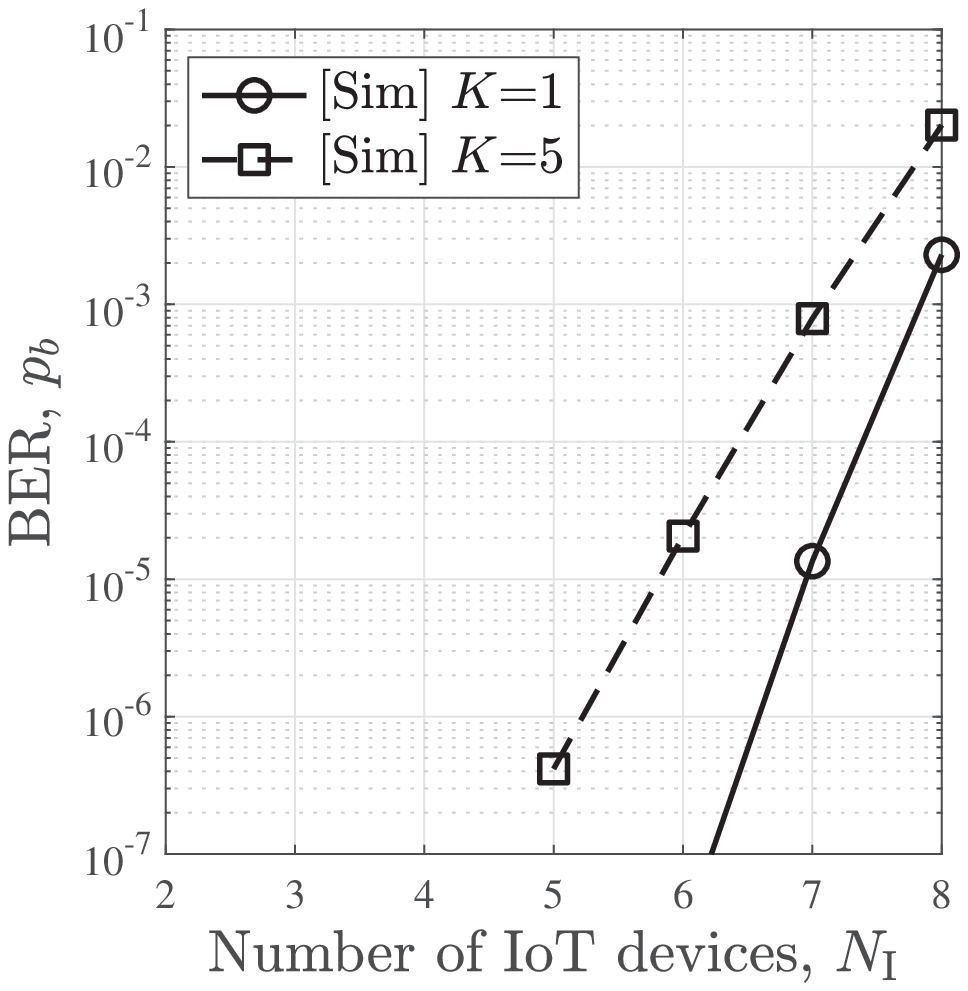}
		\caption{Bit error rate (BER)}
	\end{subfigure}	
	\caption{Collision probability and the BER performance}
	\label{fig:Fig6}
\end{figure}

Fig.~\ref{fig:Fig6} shows both a collision probability and the BER for varying the number of IoT devices, $N_\mathrm{I}$, when $M=8$ and the SNR is set to $23$~dB. 
Note that a full rank ZF matrix is unavailable when preamble collisions occur. Thus, we measure bit errors when no preamble collision occurs to exclusively investigate reliability of sending short-packets. As $N_\mathrm{I}$ increases, $p_c$ increases since the number of RA preambles is limited. The accuracy of channel estimations degrades as $N_\mathrm{I}$ increases, which consequently increases $p_b$. Increasing $K$ helps to mitigate $p_c$ but degrades $p_b$ due to the noise-rise among non-orthogonal RA preambles. Thus, there is a trade-off relationship between $p_c$ and $p_b$ with respect to $K$. 

%%%%%%%%%%%%%%%%%%%%%%%%%%%%%%%%%%%%%%%%%%%%%%%%%%%%%%%%%%%%%%%%%%%%%
%%%%%%%%%%%%%%%%%%%%%%%%%%%%%%%%%%%%%%%%%%%%%%%%%%%%%%%%%%%%%%%%%%%%%

\begin{figure}[t]
	\centering
	\includegraphics[width=8.7cm]{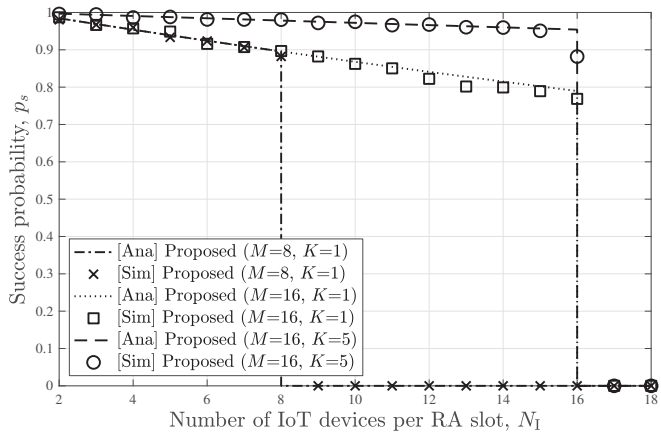}
	\caption{Success probability}
	\label{fig:Fig7}
\end{figure}

Fig.~\ref{fig:Fig7} shows a success probability for varying $N_\mathrm{I}$ when the SNR is set to 23~dB. We verify that $p_s$ highly depends on $p_c$ rather than $p_b$. Our proposed scheme can utilize multiple root sequences while avoiding a significant degradation in BER. Thus, using a few more root sequences is sufficient to significantly increase $p_s$, which enables to achieve low-latency (e.g., $6$ ms) at the same time.\footnote{1 ms (preamble transmission) + 1 ms (short-packet transmission) + 3 ms (processing time at the eNodeB) + 1 ms (acknowledgement) = 6 ms.}
If we consider the same latency constraint, then more opportunities for short-packet transmissions can be expected with our proposed scheme, compared to the conventional one. Furthermore, it is noteworthy that the number of supportable devices is limited due to the physical limitation, i.e., the number of antennas at the eNodeB.

%%%%%%%%%%%%%%%%%%%%%%%%%%%%%%%%%%%%%%%%%%%%%%%%%%%%%%%%%%%%%%%%%%%%%
%%%%%%%%%%%%%%%%%%%%%%%%%%%%%%%%%%%%%%%%%%%%%%%%%%%%%%%%%%%%%%%%%%%%%

\section{Conclusions}
In this letter, we proposed the enhanced RA with preamble-assisted short-packet transmissions in cellular IoT networks. The key feature of the proposed scheme is to acquire uplink CSI of IoT devices using the RA preambles, which can support short-packet transmissions along with the RA procedure without extra signalings. Simulation results show that mitigating the collision probability is of importance to improve the success probability of short-packet transmissions. Thus, only a few more root sequences are sufficient to newly provide an additional functionality of reliable and low-latency featured short-packet transmissions along with the RA procedure.

\bibliographystyle{IEEEtran}

\end{document}